\documentclass[10pt]{article}
\usepackage[utf8]{inputenc}
\pdfoutput=1
\usepackage{natbib}
\usepackage[english]{babel}
\usepackage{hyperref,graphicx,xspace,rotating}
\usepackage[pdftex]{color}
\usepackage{tikz}
\usepackage[labelsep=period,labelfont=bf]{caption}
\usepackage{geometry}
\usepackage{amsmath,amssymb,textcomp,url}

\begin{document}
\thispagestyle{empty}
\begin{center}
\Large Thomas Ruedas\textsuperscript{1,2}\\[5ex]
\textbf{Radioactive heat production of six geologically important nuclides}\\[5ex]
final version\\[5ex]
6 September 2017\\[10ex]
published in:\\
\textit{Geochemistry, Geophysics, Geosystems}, 18(9), pp.~3530--3541 (2017)\\[15ex]
\normalsize
\textsuperscript{1}Institute of Planetology, Westf\"alische Wilhelms-Universit\"at, M\"unster, Germany\\[5ex]
\textsuperscript{2}Institute of Planetary Research, German Aerospace Center (DLR), Berlin, Germany
\rule{0pt}{12pt}
\end{center}
\vfill
\footnotesize An edited version of this paper was published by AGU. Copyright (2017) American Geophysical Union. That version of record is available at \url{http://dx.doi.org/10.1002/2017GC006997}.
\normalsize
\newpage

\title{Radioactive heat production of six geologically important nuclides}
\author{Thomas Ruedas\thanks{Corresponding author: T. Ruedas, Institute of Planetology, Westf\"alische Wilhelms-Universit\"at, M\"unster, Germany (t.ruedas@uni-
muenster.de)}\\{\footnotesize Institute of Planetology, Westf\"alische Wilhelms-Universit\"at, M\"unster, Germany}\\{\footnotesize Institute of Planetary Research, German Aerospace Center (DLR), Berlin, Germany}}
\date{}
\maketitle
\textbf{Key points}
\begin{itemize}
\item The heat production rates of six geologically important radionuclides are reevaluated.
\item Rates differ from some older evaluations by several per cent.
\item Recent data agree to within 1\%, except for K-40.
\end{itemize}
\begin{abstract}
Heat production rates for the geologically important nuclides ${}^{26}$Al, ${}^{40}$K, ${}^{60}$Fe, ${}^{232}$Th, ${}^{235}$U, and ${}^{238}$U are calculated on the basis of recent data on atomic and nuclear properties. The revised data differ by several per cent from some older values, but indicate that more recent analyses converge toward values with an accuracy sufficient for all common geoscience applications, although some possibilities for improvement still remain, especially in the case of ${}^{40}$K and with regard to the determination of half-lives. A Python script is provided for calculating heat production (\url{https://github.com/trg818/radheat}).
\end{abstract}

\section{Introduction}
All bodies in the Solar System were endowed with certain amounts of different radioactive nuclides at the time of their formation. The six nuclides considered here appear mostly in the silicate parts of planetary bodies, except for ${}^{60}$Fe, of which a major part would have occurred in the metallic phase. The concentrations of the four nuclides still extant (${}^{40}$K, ${}^{232}$Th, ${}^{235}$U, and ${}^{238}$U) in the bulk silicate parts of terrestrial planets or asteroidal material lie on the order of tens of ppb to hundreds of ppt \citep[e.g.,][]{VanSchmus95,LoFe11,PaONe14}, but they can increase by more than one or two orders of magnitude in oceanic and terrestrial continental crust, respectively, as a consequence of their strongly incompatible behavior in the melting processes that create crustal rocks \citep{VanSchmus95,Jaup:etal15}. The shorter-lived nuclides ${}^{26}$Al and ${}^{60}$Fe seem to have had similar or somewhat lower concentrations \citep[e.g.,][]{Teng17,ElSt17}. For many of the Solar System bodies, especially the terrestrial planets or various types of asteroids, the decay of these nuclides has been, and still is, an important internal heat source that drives processes like mantle convection and provides much of the energy for the production of melt that results in volcanism at the surface. In the early evolution of the Solar System, especially ${}^{26}$Al and ${}^{60}$Fe have been considered important heat sources that facilitated extensive melting of the forming planets and their differentiation \citep{Urey55a,KoRo80}, although very recent work indicates that the concentration of ${}^{60}$Fe has always been too low to render this nuclide important as a heat source \citep{Boeh:etal17,Trap:etal17}. Models of the evolution of planetary interiors therefore rely on accurate heat production rates for the most important nuclides in order to account correctly for the magnitude and temporal change of this heat source.\par
Data on various atomic and nuclear properties and isotopic abundances are gathered frequently by different institutions, and every few years, international working groups release datasets of recommended values for many physical constants \citep{CODATA2014} as well as atomic and nuclear properties \citep[e.g.,][]{Audi:etal17,MWang:etal17,Meij:etal16a}. In this technical note, recent relevant data are gathered in order to calculate heat production rates for the six geologically most important heat-producing nuclides ${}^{26}$Al, ${}^{40}$K, ${}^{60}$Fe, ${}^{232}$Th, ${}^{235}$U, and ${}^{238}$U, as the data frequently used in the literature are getting out of date or have not always been determined correctly or accurately. This paper therefore serves the dual purpose of re-evaluating and updating radioactive heat production data as well as providing a short summary of the relevant basic physics. A Python script is included to enable the users to calculate heat production themselves using their own data.

\section{Theoretical background}\label{sect:theo}
In principle, the energy $Q$ released by radioactive decay is given by the difference between the mass of the parent, $m_P$, and the daughter nuclide(s), $m_D$, multiplied by the square of the velocity of light in vacuum, $c_0$. However, in $\beta^-$, $\beta^+$, and $\epsilon$ (electron capture, ec) decays, a part $E_\nu$ of the energy is carried away by a neutrino or antineutrino, whose interaction with matter is almost nil and which therefore does not contribute to heat production or to driving chemical processes such as radiolysis; indeed, although this paper focuses on heat, the core issue really is how much of the decay energy causes any effect in matter, and so the following considerations concerning heat energy and power apply also to chemical reactions. The formation of an (anti)neutrino in those three decay types also ensures the conservation of momentum, angular momentum, and spin, which would otherwise be violated. The energy available for heat production and the resulting power are therefore
\begin{align}
E_H&=(m_P-m_D)c_0^2-E_\nu\\
H&=E_H\lambda_{1/2},
\end{align}
where the decay rate $\lambda_{1/2}=(\ln 2)/T_{1/2}$ is determined from the half-life $T_{1/2}$. The necessity to subtract the neutrino fraction of the energy has not always been appreciated in previous evaluations of heat production, as already noted by \citet{Ca-Ro:etal09}. Even if the neutrinos were accounted for, several previous workers have made the simplifying assumption that the neutrino always carries away 2/3 of the maximum $\beta$ energy. However, this is an empirically found approximate average of the energy of the neutrino, which may over- or underestimate the actual value substantially; the longer the decay chain is, the more one may hope that the over- and underestimations cancel in the sum, but especially for short decay chains, this cannot be relied on.\par
For most radionuclides, there are several possibilities to decay even for a given type of decay, because the emission or capture of particles does not always transform the parent nucleus directly into the ground state of the daughter nucleus but can leave it in some excited state, of which there can be several. The ground state is then reached by emission of gamma or X rays. In decay types that involve neutrinos, the different intermediate states must be considered individually in order to account correctly for the energy fraction lost to the (anti)neutrino in reaching them. This is the approach that is followed in this paper.\par
While striving for accuracy in the calculation of the heat production, two simplifying assumptions are nonetheless made:
\begin{itemize}
\item Often a nuclide has several decay branches, but some of them are several orders of magnitude less probable than others. Such branches with very low probability contribute almost nothing to the heat output and are often neglected in the calculations. In particular, spontaneous fission of ${}^{232}$Th, ${}^{235}$U, and ${}^{238}$U, whose probability is less than $10^{-6}$ \citep[\textsc{Nubase2016,}][]{Audi:etal17}, is not considered at all.
\item ${}^{60}$Fe, ${}^{232}$Th, ${}^{235}$U, and ${}^{238}$U do not decay directly into stable daughters but into nuclides that are also unstable and decay further. However, the half-lives of all daughters considered here happen to be several orders of magnitude shorter than those of the initial nuclide and do not exceed a few years in most cases. The rate-limiting process for the entire decay chain and its heat production is therefore the decay of the initial nuclide, and it is assumed here that its half-life applies to the decay chain as a whole; in other words, parent and daughters are assumed to be in secular equilibrium. For the three long-lived heavy nuclides ${}^{232}$Th, ${}^{235}$U, and ${}^{238}$U, whose decay chains include ${}^{220}$Rn, ${}^{219}$Rn, and ${}^{222}$Rn, respectively, this assumption requires that none of this gas is lost. Given that the half-lives of these Rn isotopes (55.6\,s, 3.96\,s, and 3.8215\,d, respectively; \citep[\textsc{Nubase2016,}][]{Audi:etal17}) are much shorter than generally expected migration times of atoms in large volumes of solid bulk geological materials, this seems justified, but there may be special circumstances such as near-surface environments where this is not strictly the case.
\end{itemize}
In the following, the relevant decay types are shortly described.

\subsection{$\alpha$ decay}
$\alpha$ decay affects mostly nuclei with intermediate to high atomic number; the long decay series of the heavy isotopes ${}^{232}$Th, ${}^{235}$U, and ${}^{238}$U include several $\alpha$ decays. The general nuclear reaction for $\alpha$ decay from parent $P$ into daughter $D$ is
\begin{equation}
{}_Z^AP \to [{}_{Z-2}^{A-4}D]^{2-}+\alpha \to {}_{Z-2}^{A-4}D+{}_{2}^{4}\mathrm{He},
\end{equation}
where $A$ and $Z$ are the nucleon and the atomic number, respectively. As $\alpha$ decay does not involve (anti)neutrinos but is limited to two bodies, the spectrum of decay energies to the different discrete energy levels of the daughter nuclide is also discrete, and all of the decay energy is ultimately transformed into heat and can be expressed to a very good approximation as
\begin{equation}
E_{H\alpha}=(m_P-m_D-m_{^4\mathrm{He}})c_0^2,\label{eq:EHalpha}
\end{equation}
taking into account the mass of the ${}^{4}$He atom that eventually forms from the $\alpha$ particle \citep[e.g.,][]{MaGa05}; the difference in the electron binding energies between the parent and the decay products is neglected here.

\subsection{$\beta^-$ decay}
At the core of the $\beta^-$ decay is the transformation of a neutron into a proton, a high-velocity electron ($\beta^-$), and an antineutrino: $\mathrm{n}\to\mathrm{p}+\beta^-+\bar{\nu}$. The reaction is thus
\begin{equation}
_Z^AP \to [_{Z+1}^{\quad A}D]^+ +\beta^- +\bar{\nu} \to {_{Z+1}^{\quad A}D} +\bar{\nu}
\end{equation}
\citep[e.g.,][]{MaGa05}; the cation is quickly neutralized to $_{Z+1}^{\quad A}D$ by an electron from the surrounding medium. It occurs in neutron-rich nuclei, among them all of the parent isotopes considered here except ${}^{26}$Al. Contrary to the $\alpha$ decay, the $\beta^-$ does not have a discrete energy spectrum but covers an energy continuum between a characteristic maximum, or endpoint, energy $E_{\beta,\mathrm{max}}$ and zero. The reason for this is that the transition from the parent to the daughter nuclide involves the emission not of only a single particle but of two, namely the $\beta^-$ electron and the $\bar{\nu}$; the total energy difference $E_{\beta,\mathrm{max}}$ between the ground state of the parent and the state of the daughter reached in this stage of the decay can be divided between them in any proportion, as long as both contributions add up to $E_{\beta,\mathrm{max}}$. If the mean energy of the $\beta^-$ is $\langle E_\beta\rangle$, then only a fraction $X_\beta=\langle E_\beta\rangle/E_{\beta,\mathrm{max}}$ of the total energy difference of the particle emission stage is available as heat, and the total heat released is
\begin{equation}
E_{H\beta^-}=\sum\limits_i \left(\langle E_\beta\rangle_i+E_{\gamma i}\right)=
\sum\limits_i \left[\langle E_\beta\rangle_i+(m_P-m_D)c_0^2-E_{\beta i,\mathrm{max}}\right],
\end{equation}
where the sum is over all states of the daughter that are reached by the nucleus after the emission of the $\beta^-$ and $\bar{\nu}$. The $E_{\gamma i}$ indicate the energy subsequently released as electromagnetic radiation as the daughter transitions from the excited into the ground state.

\subsection{$\beta^+$ decay}
The $\beta^+$ decay is in many respects symmetric to the $\beta^-$ decay; it occurs in proton-rich nuclei, most importantly for our purposes in ${}^{26}$Al, but also, albeit with negligible intensity, in ${}^{40}$K. The underlying process here is the transformation of a proton into a neutron, a high-velocity positron ($\beta^+$), and a neutrino: $\mathrm{p}\to\mathrm{n}+\beta^+ +\nu$, and the reaction is
\begin{equation}
_Z^AP \to [_{Z-1}^{\quad A}D]^- +\beta^+ +\nu \to {_{Z-1}^{\quad A}D} +\nu
\end{equation}
\citep[e.g.,][]{MaGa05}. The anion is neutralized to $_{Z-1}^{\quad A}D$ by losing an electron to the surrounding medium; however, in addition the $\beta^+$ annihilates with an electron from the medium, producing two $\gamma$ photons with energies of $\sim 511$\,keV each that contribute to the heat output. The energy spectrum considerations are analogous to the $\beta^-$ decay, and so the total heat released can be written as
\begin{equation}
E_{H\beta^+}=\sum\limits_i \left(\langle E_\beta\rangle_i+E_{\pm i}+E_{\gamma i}\right)=
\sum\limits_i \left[\langle E_\beta\rangle_i+(2m_\mathrm{el}+m_P-m_D)c_0^2-E_{\beta i,\mathrm{max}}\right],
\end{equation}
where the sum is over all states of the daughter that are reached by the nucleus after the emission of the $\beta^+$ and $\nu$. $E_{\pm i}=2m_ec_0^2$ is the energy corresponding to the rest mass of the electron and the positron from a decay to the $i$th level that are annihilated.

\subsection{Electron capture ($\epsilon$, ec)}
This process involves the capture of an electron from one of the inner orbitals of the parent atom by the parent nucleus, leading to the reaction $\mathrm{p}+\mathrm{e}^-\to\mathrm{n}+\nu$. In terms of the final products, it is thus very similar to the $\beta^+$ decay, but with the important difference that no positron is emitted and therefore no annihilation takes place:
\begin{equation}
_Z^AP \to {_{Z-1}^{\quad A}D} +\nu
\end{equation}
\citep[e.g.,][]{MaGa05}. An energy level of the daughter nucleus is reached by the emission of a neutrino only, and as a consequence, all heat release from ec is due to the subsequent de-excitation of the daughter, i.e.,
\begin{equation}
E_{H\epsilon}=\sum\limits_i E_{\gamma i}.
\end{equation}
This process also occurs in proton-rich nuclei, specifically in ${}^{26}$Al and ${}^{40}$K.

\section{Results}
In the calculations, half-lives and isotope masses are taken from the \textsc{Nubase}2016 and \textsc{Ame}2016 evaluations \citep{Audi:etal17,MWang:etal17}, and mean atomic masses and isotopic abundances are taken from the corresponding IUPAC 2013 reports \citep{Meij:etal16a,Meij:etal16b}, unless otherwise stated; physical constants are from CODATA2014 \citep{CODATA2014} (Table~\ref{tab:const}). The nuclide-specific decay data (energies and intensities) were obtained from the NuDat v.2.7$\beta$ (as retrieved in early July 2017) website (\url{www.nndc.bnl.gov/nudat2}) of the National Nuclear Data Center (NNDC) of the Brookhaven National Laboratory, USA, and the Recommended Data site of the Laboratoire National Henri Becquerel, Saclay, France (\url{www.nucleide.org/DDEP_WG/DDEPdata.htm}); in general, the newest available data were used. The main results are listed in Table~\ref{tab:res}.
\begin{table}
\centering
\caption{Physical constants and unit conversion factors used in the calculations\label{tab:const}}
\begin{tabular}{lccl}\hline
Constant&Symbol&Value&Reference\\\hline
Speed of light in vacuum&$c_0$&299792458\,m/s&CODATA2014\\
Atomic mass unit (a.m.u.)&$u$&$1.660539040\times 10^{-27}$\,kg&CODATA2014\\
Rest mass of electron&$m_\mathrm{el}$&$9.10938356\times 10^{-31}$\,kg&CODATA2014\\
Rest mass of ${}^{4}$He&$m_{^4\mathrm{He}}$&4.00260325413\,a.m.u.&\textsc{Ame}2016\\
Elementary charge&$e$&$1.6021766208\times 10^{-19}$\,C&CODATA2014\\
Year&--&31556926\,s&\textsc{Nubase}2016\\
Year&--&365.2422 days&\textsc{Nubase}2016\\\hline
\end{tabular}
\end{table}
\begin{sidewaystable}
\centering
\caption{Atomic mass ($u$), half-life ($T_{1/2}$), present-day fractional abundance ($X_\mathrm{iso}$), total and heat-effective decay energy per atom ($Q$, $E_H$), and specific heat production ($H$) of ${}^{26}$Al, ${}^{40}$K, ${}^{60}$Fe, ${}^{232}$Th, ${}^{235}$U, and ${}^{238}$U. The decay energies of ${}^{40}$K are the weighted means of the two principal decay modes. For multi-step decays, only the total values are given. The elemental specific heat productions for K and U are determined according to Eq.~\ref{eq:Helm} as described in Sect.~\ref{sect:power}.\label{tab:res}}
\begin{tabular}{lcccccccc}\hline
&$u$ (a.m.u.)&$T_{1/2}$ (My)&$X_\mathrm{iso}$&\multicolumn{2}{c}{$Q$ (keV, J)}&\multicolumn{2}{c}{$E_H$ (keV, J)}&$H$ (W/kg)\\\hline
${}^{26}$Al&25.986891863&0.717&0&4004.393&$6.416\times 10^{-13}$&3150.155&$5.047\times 10^{-13}$&0.3583\\
${}^{40}$K&39.963998166&1248&$1.1668\times 10^{-4}$&1331.637&$2.134\times 10^{-13}$&676.863&$1.084\times 10^{-13}$&$2.8761\times 10^{-5}$\\
K&39.0983&&&&&&&$3.4302\times 10^{-9}$\\
${}^{60}$Fe&59.934070411&2.62&0&3060.102&$4.903\times 10^{-13}$&2710.251&$4.342\times 10^{-13}$&$3.6579\times 10^{-2}$\\
${}^{232}$Th&232.038053689&14000&1&42645.892&$6.833\times 10^{-12}$&40418.037&$6.476\times 10^{-12}$&$2.6368\times 10^{-5}$\\
${}^{235}$U&235.043928190&704&0.0072045&46396.500&$7.434\times 10^{-12}$&44379.817&$7.110\times 10^{-12}$&$5.68402\times 10^{-4}$\\
${}^{238}$U&238.050786996&4468&0.9927955\textsuperscript{*}&51694.046&$8.282\times 10^{-12}$&47650.476&$7.634\times 10^{-12}$&$9.4946\times 10^{-5}$\textsuperscript{*}\\
U&238.02891\textsuperscript{**}&&&&&&&$9.8314\times 10^{-5}$\\\hline
\end{tabular}\\[1ex]
{\footnotesize \textsuperscript{*} Includes the fraction/contribution of ${}^{234}$U ($X_\mathrm{iso}=5.4\times 10^{-5}$).}\rule{11.23cm}{0pt}\\[1ex]
{\footnotesize \textsuperscript{**} The atomic mass of ${}^{234}$U is 234.040950370 a.m.u.}\rule{12.9cm}{0pt}
\end{sidewaystable}
\subsection{${}^{26}$Al}
${}^{26}$Al was probably the most important heat source during the early formation stages of planets and is now extinct in planetary bodies. As Al is a lithophile element and an essential constituent of several major minerals, this heat source will have been significant mostly in the rocky parts of planet-forming materials. ${}^{26}$Al decays with a half-life of $\sim 717$\,ky into the stable ${}^{26}$Mg, either by $\beta^+$ decay with a probability $I$ of 0.8173 or by electron capture with a probability of 0.1827 \citep[NuDat v.2.7$\beta$,][]{BaHu16}:\\
\begin{tikzpicture}[scale=0.9]
\draw (0,0) node {${}^{26}$Al};
\draw [->] (0.6,0.2) arc (110:70:7ex) node[pos=0.5,above] {$\beta^+$};
\draw [->] (0.6,-0.2) arc (-110:-70:7ex) node[pos=0.5,below] {$\epsilon$};
\draw (2,0) node {${}^{26}$Mg};
\end{tikzpicture}\\
Almost one quarter of the total decay energy is carried off by neutrinos and does not contribute to heat production.
\subsection{${}^{40}$K}
${}^{40}$K is the most abundant of the four still extant nuclides considered here and is of special interest because it is assumed to contribute most of the radiogenic heat of the core, as it is the one of the four long-lived nuclides that is thought to partition most easily into a metallic melt \citep[e.g.,][]{GeWo02,Murt:etal03,Bouh:etal07b,Corg:etal07,KWata:etal14}; nonetheless, K is rather lithophile, and so the actual K content of the core, which depends strongly on the particular composition of the core alloy, is probably only on the order of a few tens of ppm. As a consequence, most of it resides in the silicate part of a planet. Being an incompatible element, it strongly partitions into silicate melt and is particularly strongly concentrated in the rock types of the continental crust of the Earth, which on average reaches K contents of a few per cent \citep{VanSchmus95,Jaup:etal15}.\par
The present-day fraction of ${}^{40}$K in K is $1.1668\times 10^{-4}$ \citep{Naum:etal13}. It has two important decay branches, both of which produce stable daughters:\\
\begin{tikzpicture}[scale=0.9]
\draw (0,0) node {${}^{40}$K};
\draw [->] (0.7,0) -- (1.4,0.4) node[pos=0.4,above] {$\beta^-$};
\draw [->] (0.7,0) -- (1.4,-0.4) node[pos=0.3,below] {$\epsilon$};
\draw (2,0.4) node {${}^{40}$Ca};
\draw (2,-0.4) node {${}^{40}$Ar};
\end{tikzpicture}\\
The most likely ($I=0.8928$) decay is a $\beta^-$ decay directly to the ground state of ${}^{40}$Ca, the other ($I=0.1072$) produces ${}^{40}$Ar by electron capture, as does the third, very low-probability $\beta^+$ branch \citep[NuDat v.2.7$\beta$,][]{JChen17}. The half-life is $\sim1.248$\,Gy \citep[\textsc{Nubase}2016,][]{Audi:etal17}. The evaluations from both NuDat \citep{JChen17} and \url{www.nucleide.org} \citep{MoHe12} were found to underestimate $\langle E_\beta\rangle$ for the $\beta^-$ branch significantly, and so this value was determined to be 583.55\,keV with the program BetaShape, which overcomes the limitations responsible for the underestimates \citep[and pers. comm., 2017]{Mougeot15}.
\subsection{${}^{60}$Fe}
The other major radiogenic heat source during the earliest part of Solar System history may have been ${}^{60}$Fe ($T_{1/2}\approx 2.62$\,My), which will have occurred in both the rocky and the metallic parts of forming planetary bodies. However, its importance depends on its actual concentration, for which reported measurements vary by a factor of 60; ongoing work suggests that some older measurements have yielded too high values and that its role may thus be overestimated \citep{Boeh:etal17,Trap:etal17}. ${}^{60}$Fe transmutes by two $\beta^-$ decays via ${}^{60}$Co ($T_{1/2}\approx 5.2712$\,yr) to ${}^{60}$Ni \citep[NuDat v.2.7$\beta$,][]{BrTu13b}:\\
\begin{tikzpicture}[scale=0.9]
\draw (0,0) node {${}^{60}$Fe};
\draw [->] (0.7,0) -- (1.4,0) node[pos=0.5,below] {$\beta^-$};
\draw (2,0) node {${}^{60}$Co};
\draw [->] (2.7,0) -- (3.4,0) node[pos=0.5,below] {$\beta^-$};
\draw (4,0) node {${}^{60}$Ni};
\end{tikzpicture}\\
and is now extinct in planetary interiors. As the half-life of the intermediate step is more than five orders of magnitude shorter than that of ${}^{60}$Fe, the $T_{1/2}$ of the latter is assumed to apply to the entire chain. However, it is the second decay step that contributes more than 95\% of the total heat.
\subsection{${}^{232}$Th}
Of the six radionuclides considered, ${}^{232}$Th is the one with the longest half-life ($T_{1/2}\approx 14$\,Gy), and it is the only isotope of Th that exists today in appreciable amounts. Virtually all of it is thought to reside in the silicate parts of the planets, especially in the crust, as Th is not siderophile and behaves very incompatibly upon silicate melting \citep{BJWoBl14}. It has a long decay chain consisting of several $\alpha$ and $\beta^-$ decays:\\
\begin{tikzpicture}[scale=0.9]
\draw (0,0) node {${}^{232}$Th};
\draw [->] (0.7,0) -- (1.4,0) node[pos=0.5,below] {$\alpha$};
\draw (2,0) node {${}^{228}$Ra};
\draw [->] (2.7,0) -- (3.4,0) node[pos=0.5,below] {$\beta^-$};
\draw (4,0) node {${}^{228}$Ac};
\draw [->] (4.7,0) -- (5.4,0) node[pos=0.5,below] {$\beta^-$};
\draw (6,0) node {${}^{228}$Th};
\draw [->] (6.7,0) -- (7.4,0) node[pos=0.5,below] {$\alpha$};
\draw (8,0) node {${}^{224}$Ra};
\draw [->] (8.7,0) -- (9.4,0) node[pos=0.5,below] {$\alpha$};
\draw (10,0) node {${}^{220}$Rn};
\draw [->] (10.7,0) -- (11.4,0) node[pos=0.5,below] {$\alpha$};
\draw [->] (2.7,-1.5) -- (3.4,-1.5) node[pos=0.5,below] {$\alpha$};
\draw (4,-1.5) node {${}^{216}$Po};
\draw [->] (4.7,-1.5) -- (5.4,-1.5) node[pos=0.5,below] {$\alpha$};
\draw (6,-1.5) node {${}^{212}$Pb};
\draw [->] (6.7,-1.5) -- (7.4,-1.5) node[pos=0.5,below] {$\beta^-$};
\draw (8,-1.5) node {${}^{212}$Bi};
\draw [->] (8.7,-1.5) -- (9.4,-1.1) node[pos=0.4,above] {$\beta^-$};
\draw [->] (8.7,-1.5) -- (9.4,-1.9) node[pos=0.4,below] {$\alpha$};
\draw (10,-1.1) node {${}^{212}$Po};
\draw [->] (10.7,-1.1) -- (11.4,-1.45) node[pos=0.6,above] {$\alpha$};
\draw (10,-1.9) node {${}^{208}$Tl};
\draw [->] (10.7,-1.9) -- (11.4,-1.55) node[pos=0.6,below] {$\beta^-$};
\draw (12,-1.5) node {${}^{208}$Pb};
\end{tikzpicture}\\
but as the half-lives of all unstable daughters are between 9 and 18 orders of magnitude shorter, $T_{1/2}=14$\,Gy is used for all rate calculations. While heat production calculation from the $\alpha$ decays is straightforward and requires only the masses of the parent and daughter (Eq.~\ref{eq:EHalpha}), care must be taken with the $\beta^-$ decays of ${}^{228}$Ra \citep{Luca09a,Luca12}, ${}^{228}$Ac \citep[NuDat v.2.7$\beta$,][]{Abusaleem14}, ${}^{212}$Pb \citep{ALNichols11h,ALNichols11g}, ${}^{212}$Bi \citep{ALNichols11j,ALNichols11i}, and ${}^{208}$Tl \citep{ALNichols10,ALNichols16}, all of which feed several different excited energy levels of their respective daughters. Moreover, ${}^{212}$Bi has two decay branches: it can decay via the $\alpha$ branch ($I=0.3593$) to ${}^{208}$Tl or via the $\beta^-$ branch ($I=0.6407$) to ${}^{212}$Po \citep{ALNichols11i}, and the decay energies of the daughters have to be weighted accordingly; note that direct $\alpha$ decays from some excited transitional states of ${}^{212}$Po to ${}^{208}$Pb are not treated separately, because they make no difference in the total energy balance. Apart from the final stable daughter ${}^{208}$Pb, each decay of ${}^{232}$Th eventually also produces 6 atoms of ${}^{4}$He.
\subsection{${}^{235}$U}
Similar to Th, U is considered essentially lithophile and hence resides in the silicate parts of planets, and as it is also highly incompatible in most conditions, it is strongly concentrated in planetary crusts; however, recent work suggests that under very reducing conditions it may also partition to some extent into a metallic phase and has thus been suggested to function as a minor heat source in planetary cores \citep{WoWo15,WoWo17}. Of the two major uranium isotopes, ${}^{238}$U and ${}^{235}$U, the latter has a lower abundance and shorter half-life ($T_{1/2}=704$\,My \citep[e.g.,][]{XHuWa14}. Its decay chain is even longer and more complicated than that of ${}^{232}$Th and yields ${}^{207}$Pb as well as 7 atoms of ${}^{4}$He as final products:\\
\begin{tikzpicture}[scale=0.9]
\draw (0,0) node {${}^{235}$U};
\draw [->] (0.7,0) -- (1.4,0) node[pos=0.5,below] {$\alpha$};
\draw (2,0) node {${}^{231}$Th};
\draw [->] (2.7,0) -- (3.4,0) node[pos=0.5,below] {$\beta^-$};
\draw (4,0) node {${}^{231}$Pa};
\draw [->] (4.7,0) -- (5.4,0) node[pos=0.5,below] {$\alpha$};
\draw (6,0) node {${}^{227}$Ac};
\draw [->] (6.7,0) -- (7.4,0.4) node[pos=0.4,above] {$\beta^-$};
\draw [->] (6.7,0) -- (7.4,-0.4) node[pos=0.4,below] {$\alpha$};
\draw (8,0.4) node {${}^{227}$Th};
\draw [->] (8.7,0.4) -- (9.4,0.05) node[pos=0.6,above] {$\alpha$};
\draw (10,0) node {${}^{223}$Ra};
\draw (8,-0.4) node {${}^{223}$Fr};
\draw [->] (8.7,-0.4) -- (9.4,-0.05) node[pos=1,below] {$\beta^-$};
\draw [->] (8.7,-0.4) -- (9.4,-0.8) node[pos=0.5,below] {$\alpha$};
\draw (10,-0.8) node {${}^{219}$At};
\draw [->] (10.7,-0.8) -- (11.4,-0.8) node[pos=0.5,below] {$\alpha$};
\draw (12,-0.8) node {\dots};
\draw [->] (10.7,0) -- (11.4,0) node[pos=0.5,below] {$\alpha$};
\draw (12,0) node {${}^{219}$Rn};
\draw [->] (12.7,0) -- (13.4,0) node[pos=0.5,below] {$\alpha$};
\draw [->] (3.7,-2) -- (4.4,-2) node[pos=0.5,below] {$\alpha$};
\draw (5,-2) node {${}^{215}$Po};
\draw [->] (5.7,-2) -- (6.4,-2) node[pos=0.5,below] {$\alpha$};
\draw (7,-2) node {${}^{211}$Pb};
\draw [->] (7.7,-2) -- (8.4,-2) node[pos=0.5,below] {$\beta^-$};
\draw (9,-2) node {${}^{211}$Bi};
\draw [->] (9.7,-2) -- (10.4,-1.6) node[pos=0.4,above] {$\beta^-$};
\draw [->] (9.7,-2) -- (10.4,-2.4) node[pos=0.4,below] {$\alpha$};
\draw (11,-1.6) node {${}^{211}$Po};
\draw [->] (11.7,-1.6) -- (12.4,-1.95) node[pos=0.6,above] {$\alpha$};
\draw (11,-2.4) node {${}^{207}$Tl};
\draw [->] (11.7,-2.4) -- (12.4,-2.05) node[pos=0.6,below] {$\beta^-$};
\draw (13,-2) node {${}^{207}$Pb};
\end{tikzpicture}\\
The decay branch from ${}^{223}$Fr to ${}^{219}$At has a very low probability, and therefore the subsequent steps of that branch are omitted from the calculation; they merge with the main decay path at different points. With regard to the energy balance, the $\beta^-$ decays of ${}^{231}$Th \citep[NuDat v.2.7$\beta$,]{BrTu13a}, ${}^{227}$Ac \citep[NuDat v.2.7$\beta$,][]{Kond:etal16}, ${}^{223}$Fr \citep{XHuWa09,XHuWa12}, ${}^{211}$Pb, ${}^{211}$Bi \citep[NuDat v.2.7$\beta$,][]{BSing:etal13}, and ${}^{207}$Tl \citep[NuDat v.2.7$\beta$,][]{KoLa11} require attention because of the neutrino component. Very low-probability $\beta^-$ decays of ${}^{215}$Po to ${}^{215}$At and of ${}^{207}$Tl$^\mathrm{m}$ to ${}^{207}$Pb (not shown) are neglected.
\subsection{${}^{238}$U}
The other, much more abundant isotope ${}^{238}$U has a half-life close to the age of the Solar System ($T_{1/2}=4468$\,My \citep[\textsc{Nubase}2016,][]{Audi:etal17}) and a similarly long and complex decay chain. It yields ${}^{206}$Pb as well as 8 atoms of ${}^{4}$He as final products:\\
\begin{tikzpicture}[scale=0.9]
\draw (0,0) node {${}^{238}$U};
\draw [->] (0.7,0) -- (1.4,0) node[pos=0.5,below] {$\alpha$};
\draw (2,0) node {${}^{234}$Th};
\draw [->] (2.7,0) -- (3.2,0) node[pos=0.5,below] {$\beta^-$};
\draw (4,0) node {${}^{234}$Pa\textsuperscript{m}};
\draw [->] (4.7,0) -- (5.4,0) node[pos=0.5,below] {$\beta^-$};
\draw [->] (4,-0.3) arc (190:230:5ex) node[pos=0.3,below left] {IT};
\draw (5,-1.1) node {${}^{234}$Pa};
\draw [<-] (6,-0.3) arc (340:300:5ex) node[pos=0.2,below right] {$\beta^-$};
\draw (6,0) node {${}^{234}$U};
\draw [->] (6.7,0) -- (7.4,0) node[pos=0.4,below] {$\alpha$};
\draw (8,0) node {${}^{230}$Th};
\draw [->] (8.7,0) -- (9.4,0) node[pos=0.6,below] {$\alpha$};
\draw (10,0) node {${}^{226}$Ra};
\draw [->] (10.7,0) -- (11.4,0) node[pos=0.5,below] {$\alpha$};
\draw (12,0) node {${}^{222}$Rn};
\draw [->] (12.7,0) -- (13.4,0) node[pos=0.5,below] {$\alpha$};
\draw [->] (-0.3,-3) -- (0.4,-3) node[pos=0.5,below] {$\alpha$};
\draw (1,-3) node {${}^{218}$Po};
\draw [->] (1.7,-3) -- (2.4,-2.6) node[pos=0.4,above] {$\beta^-$};
\draw [->] (1.7,-3) -- (2.4,-3.4) node[pos=0.4,below] {$\alpha$};
\draw (3,-2.6) node {${}^{218}$At};
\draw [->] (3.7,-2.6) -- (4.4,-2.95) node[pos=0.3,below] {$\alpha$};
\draw [->] (3.7,-2.6) -- (4.4,-2.2) node[pos=0.6,above] {$\beta^-$};
\draw (5,-2.2) node {\dots};
\draw (3,-3.4) node {${}^{214}$Pb};
\draw [->] (3.7,-3.4) -- (4.4,-3.05) node[pos=0.6,below] {$\beta^-$};
\draw [->] (3.7,-3.4) -- (4.4,-3.05) node[pos=0.6,below] {$\beta^-$};
\draw (5,-3) node {${}^{214}$Bi};
\draw [->] (5.7,-3) -- (6.4,-2.6) node[pos=0.4,above] {$\beta^-$};
\draw [->] (5.7,-3) -- (6.4,-3.4) node[pos=0.4,below] {$\alpha$};
\draw (7,-2.6) node {${}^{214}$Po};
\draw [->] (7.7,-2.6) -- (8.4,-2.95) node[pos=0.6,above] {$\alpha$};
\draw (7,-3.4) node {${}^{210}$Tl};
\draw [->] (7.7,-3.4) -- (8.4,-3.05) node[pos=0.6,below] {$\beta^-$};
\draw (9,-3) node {${}^{210}$Pb};
\draw [->] (9.7,-3) -- (10.4,-3) node[pos=0.5,above] {$\beta^-$};
\draw (11,-3) node {${}^{210}$Bi};
\draw [->] (11.7,-3) -- (12.4,-3) node[pos=0.5,above] {$\beta^-$};
\draw (13,-3) node {${}^{210}$Po};
\draw [->] (13.7,-3) -- (14.4,-3) node[pos=0.5,above] {$\alpha$};
\draw (15,-3) node {${}^{206}$Pb};
\end{tikzpicture}\\
The isomeric transition (IT) of ${}^{234}$Pa was treated separately because of the slightly different $\beta^-$ energies, but the very low-probability $\beta^-$ branch of ${}^{218}$At and two similar branches of ${}^{210}$Pb to ${}^{206}$Hg and of ${}^{210}$Bi to ${}^{206}$Tl (not shown) are not included. Energy is lost by neutrinos in the $\beta^-$ decays of ${}^{234}$Th \citep{Luca09b,Luca10}, ${}^{234}$Pa$^\mathrm{m}$ and ${}^{234}$Pa \citep{XHuWa11d,XHuWa11b,XHuWa11c,XHuWa11a}, ${}^{218}$Po \citep{ChBe07a,ChBe10a}, ${}^{214}$Pb \citep{ChBe10c,ChBe10b}, ${}^{214}$Bi \citep[NuDat v.2.7$\beta$,][]{SCWu09}, ${}^{210}$Tl, ${}^{210}$Pb \citep[NuDat v.2.7$\beta$,][]{Basunia14}, and ${}^{210}$Bi \citep{Chis:etal14b,Chis:etal14a}.
\subsection{Bulk element power}\label{sect:power}
The decay calculation yields the power per mass of the isotope under consideration, but it is often convenient to know the power per mass of the element, which usually has several isotopes with different total decay heat energies (if they are radioactive). This value changes with time as the proportions of the different isotopes in the isotopic composition of the element change due to their different decay rates. The values for extant isotopes in Table~\ref{tab:res} are present-day values based on observed isotopic abundances and masses. The specific power of the bulk element is
\begin{equation}
H_\mathrm{elm}=\frac{1}{\bar{u}}\sum H_i X_i u_i,\label{eq:Helm}
\end{equation}
where $\bar{u}=\sum X_i u_i$ indicates the mean value of the atomic mass for the element, and the sum is over all isotopes. Note that in elements with radioactive isotopes, this value changes with time as a function of the half-lives of all isotopes.\par
It must be ensured that $\bar{u}$ and $\sum X_i u_i$ are consistent with each other within the accuracy of the values for $\bar{u}$ from the IUPAC 2013 reports and for the $u_i$ from the \textsc{Ame}2016 evaluations. This is the case if the values can be taken from one self-consistent database, in particular if one is interested in present-day values and chooses the isotopic abundances from \citet{Meij:etal16b}; for Th, ${}^{232}$Th is the only isotope with significant abundance, and so its power per mass of isotope can be used for the bulk element. However, if data for the abundances are from different sources, additional considerations are necessary to guarantee internal consistency.\par
One such case are the extinct isotopes ${}^{26}$Al and ${}^{60}$Fe, for which abundances must be set. Assuming that they are the only radioactive isotopes of their respective elements, one can take the tabulated mean atomic mass as the mean of the stable isotopes, $\bar{u}_\mathrm{st}$. Then the mean of the element in presence of the radionuclide is constrained by the closure condition for the abundances $X_i$, i.e., $\sum X_i=1$, to be
\begin{equation}
\bar{u}=(1-X_\mathrm{rad})\bar{u}_\mathrm{st}+X_\mathrm{rad}u_\mathrm{rad},\label{eq:ubar}
\end{equation}
where the subscript ``rad'' indicates the radioactive nuclide.\par
The isotopic fraction of ${}^{40}$K and the ratio $r_{40}=X_{40}/X_{39}$ were recently redetermined by \citet{Naum:etal13}. The fraction was found to be the same as the four decades old one given in \citet{Meij:etal16b} within error, but for consistency we recalculate the (present-day) mean atomic mass using the best estimates from \citet{Naum:etal13}, $X_{40}=1.1668\times 10^{-4}$ and $r_{40}=1.25116\times 10^{-4}$, and the tabulated masses of the three relevant isotopes, of which two are stable:
\begin{equation}
\bar{u}_\mathrm{K}=\left(u_{40}-u_{41}+\frac{u_{39}-u_{41}}{r_{40}}\right)X_{40}+u_{41}.
\end{equation}
The result agrees with the IUPAC value if rounded to the same accuracy (cf. Table~\ref{tab:res}). If $X_{40}$ is set to a different value, the corresponding $r_{40}$ becomes obsolete, and an additional condition must be made instead. Here this is the assumption that $r_{41}=X_{41}/X_{39}$ is constant and given by the IUPAC 2013 value. Then the mean atomic mass of the stable isotopes is given by
\begin{equation}
\bar{u}_\mathrm{st}=\frac{u_{39}+r_{41}u_{41}}{1+r_{41}}
\end{equation}
and is inserted into Eq.~\ref{eq:ubar}.\par
For U, we assume the existence of three radioactive isotopes and use the recent reevaluation of the ratio $r_5$ of ${}^{238}$U to ${}^{235}$U by \citet{Gold:etal15}, which gave $r_5=137.794$, slightly less than the value of 137.804 derived from the IUPAC abundances, which were derived excluding meteoritic materials; nonetheless, the resulting mean atomic masses are identical at the level of accuracy of the IUPAC value. In order to determine the isotopic fractions, an additional assumption is necessary to construct a unique solution. The choice made here is to set the ratio $r_4$ of ${}^{238}$U to ${}^{234}$U to the value given by the IUPAC data, based on the assumption that all ${}^{234}$U, which appears only in minute amounts, always stems from ${}^{238}$U, of whose decay chain it is a part, and that the relative amounts of these two coupled isotopes are consistent with each other; this implies that $r_4$ is always the same, even if a different $r_5$ is used. With $r_4=18384.111$ and the closure condition $\sum X_i=1$, we have
\begin{align}
X_{238}&=\left(\frac{1}{r_4}+\frac{1}{r_5}+1\right)^{-1}\\
\bar{u}_\mathrm{U}&=X_{234}u_{234}+X_{235}u_{235}+X_{238}u_{238}=
\left(\frac{u_{234}}{r_4}+\frac{u_{235}}{r_5}+u_{238}\right)X_{238}.
\end{align}
For the purpose of calculating bulk element power we treat ${}^{234}$U together with ${}^{238}$U; the value for $X_{238}$ in Table~\ref{tab:res} thus is in fact $X_{234}+X_{238}$. 

\section{Discussion}
\begin{figure}
\includegraphics[width=\textwidth]{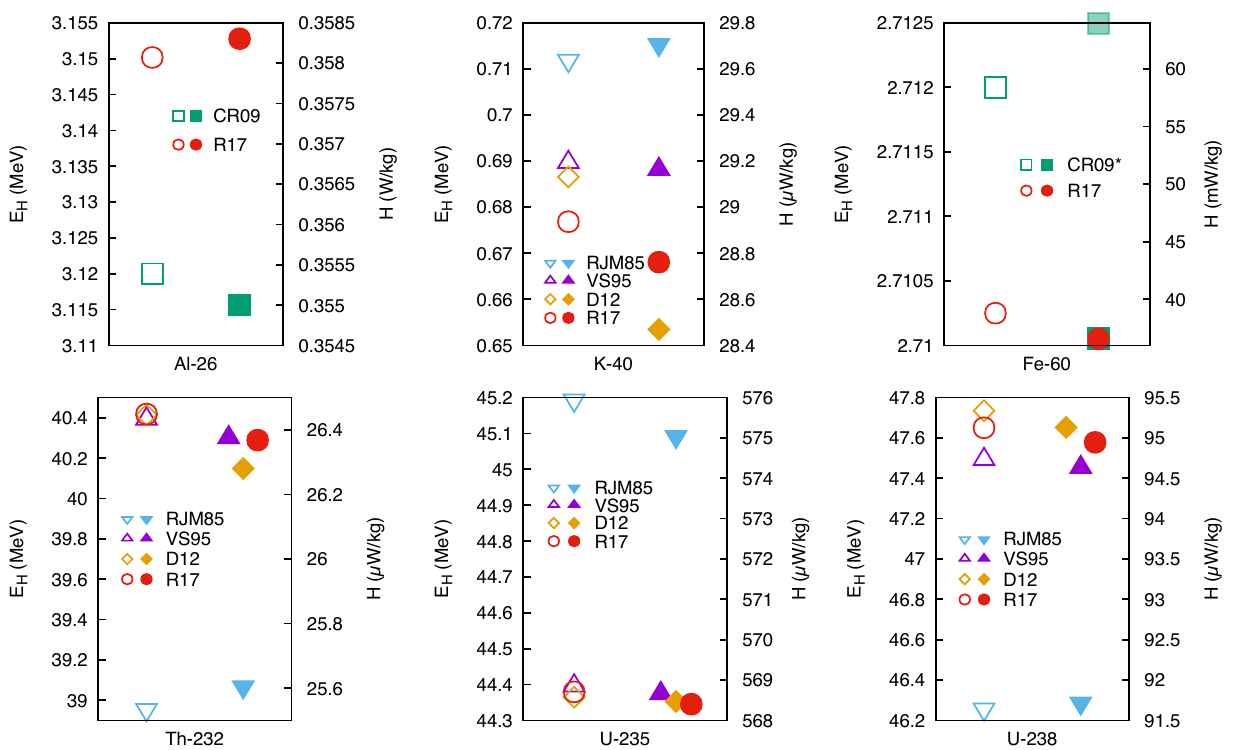}
\caption{Decay heat energy (left axis and open symbols) and specific power (right axis and solid symbols) of the radioactive decay of ${}^{26}$Al, ${}^{40}$K, ${}^{60}$Fe, ${}^{232}$Th, ${}^{235}$U, and ${}^{238}$U, as determined by \citet[RJM85]{Rybach85}, \citet[VS95]{VanSchmus95}, \citet[CR09]{Ca-Ro:etal09}, \citet[D12]{Dye12}, and in this study (R17). For ${}^{60}$Fe, CR09 did not calculate the power; therefore, two power values have been derived from their heat energy: one with the old \textsc{Nubase} value from 2003 (shown in a paler shade of green) and one with the currently accepted value used here.\label{fig:EH}}
\end{figure}

A comparison with previous evaluations shows to which extent the detailed computation and the use of up-to-date datasets improve the values of the calculated decay heat energy and the specific power. Fig.~\ref{fig:EH} summarizes the results of the following comparison.\par
${}^{26}$Al and ${}^{60}$Fe are of interest mostly in studies of planetary formation and the earliest stages of planetary evolution. The most recent and reliable determination of heat production rates of these nuclides known to me was made by \citet{Ca-Ro:etal09}. Compared to their values, the decay heat energy and power found in this study are 1\% higher in the case of ${}^{26}$Al, and the decay heat energy for ${}^{60}$Fe is practically identical; the results are hence in very good agreement. \citet{Ca-Ro:etal09} did not calculate a power for ${}^{60}$Fe; assuming that they would have used the A\textsc{me} and N\textsc{ubase} evaluations from 2003 \citep{Audi:etal03b,Audi:etal03a}, which were the most recent ones at their time, they would have arrived at a power of 63.93\,mW/kg, i.e., almost 75\% more than the value calculated here, because the half-life given in the old \textsc{Nubase} dataset is 1.5\,My. Almost all of the difference is therefore due to the change in the assigned $T_{1/2}$ (Fig.~\ref{fig:EH}, top right panel). There have been very substantial revisions of the half-life of ${}^{60}$Fe in recent years, and older power calculations for this nuclide may therefore be seriously in error even if the decay heat they used was accurate.\par
The four longer-lived nuclides are of broader relevance in the geosciences. Three reference works that provide decay energy and power information are the review paper by \citet{VanSchmus95}, which seems to be based on original calculations, and the tabulations of \citet[Table~1]{JaMa14} and \citet[Table~10]{Jaup:etal15}; the latter two, however, merely quote values from older work by \citet{Rybach85}, which in turn is based on work by the same author from the 1970s, and from a recent analysis by \citet{Dye12}, respectively. Rybach, and except for ${}^{40}$K also \citet{VanSchmus95}, make the assumption that the neutrino always carries away $2E_{\beta,\mathrm{max}}/3$, which is a relatively inaccurate simplification, as discussed above. It happens to work well for the long decay chains of Th and U, where the deviations largely cancel out in the sum of all $\beta^-$ decays involved, although the difference for individual steps in the decay chains from measured values is often several percent; for the short decay sequences of ${}^{40}$K and ${}^{60}$Fe, however, it is more than 10\% and 3\% off, respectively. By contrast, \citet{Dye12} uses actual decay spectra in a spirit similar to that of this study.\par
The decay heat energies determined in this study differ by less than 0.5\% for ${}^{232}$Th, ${}^{235}$U, and ${}^{238}$U from those by \citet{VanSchmus95} but by almost 1--3.8\% from most of those by \citet{Rybach85}. The differences in the power calculations are similar. The agreement with the values by \citet{Dye12} is excellent for these three nuclides, with differences being at most a few per mil for both heat energy and power.\par
For ${}^{40}$K, the discrepancies are a bit larger, as the decay heat energy and power computed by \citet{VanSchmus95} is 1.9 and 1.4\% higher and the values from \citet{Dye12} are higher by 1.4 and lower by 1\%, respectively. The decay heat energy and power given by \citet{Rybach85} and listed by \citet{JaMa14} are 5.1 and 3.3\% higher than those from this study, respectively. The reason for this is the oversimplification concerning the neutrino energy made in that paper as well as the fact that apparently the electron capture branch was not taken into account at all. Therefore, these values must be dismissed. The discrepancy between the decay heat determined by \citet{VanSchmus95} and the similarly high value by \citet{Dye12} on the one hand and the value from this study on the other hand is mostly due to the difference in the mean $\beta$ energy used for the calculation. \citet{VanSchmus95} used 598\,keV, and the value used by \citet{Dye12} seems to have been almost as high, although his calculation is not presented in sufficient detail to be sure, whereas the value determined with BetaShape used here is 584\,keV. In the power calculations, these high energy estimates are (over)compensated by the use of longer half-lives used in the older studies (see Fig.~\ref{fig:EH}, top center).\par
The fact that the differences between the results from this study and those by previous workers are not the same for the decay heat energy and the power  highlights one particular source of uncertainty in addition to errors and uncertainties in the energy calculation, namely the uncertainty in the determination of the half-life $T_{1/2}$. The comparatively short half-lives of the intermediate steps in the decay chains considered here are not a matter of concern in this context because of the assumption of secular equilibrium; however, a potential problem lies in the dominant half-lives of the parent. The survey of half-life measurements by \citet{BoHa14} includes some of the isotopes under consideration here and highlights some problems with the generally accepted half-lives. For instance, they point out that the accepted half-lives were determined using measurements most of which are several decades old and thus do not benefit from recent technological advances. Furthermore, the number of available measurements is very small in several cases; in particular, the accepted half-life of ${}^{238}$U, which has a crucial role in half-life evaluations, is based on one single study \citep{Jaff:etal71}, as also discussed by \citet{IMVill:etal16}. \citet{BoHa14} determined half-lives of 14.13\,Gy and 702.5\,My for ${}^{232}$Th and ${}^{235}$U, respectively, which would result in specific heat productions of $2.6126\times 10^{-5}$ W/kg for ${}^{232}$Th and of $5.69616\times 10^{-4}$ W/kg for ${}^{235}$U, other parameters being equal; this is about 1\% less and 0.2\% more, respectively, than the values in Table~\ref{tab:res}. For these nuclides, the uncertainty introduced by the half-lives is thus of approximately the same magnitude as the differences related to the decay energy between this study and the calculations by \citet{VanSchmus95}, who used the same half-life values for these two nuclides. In summary, even though the uncertainty from half-lives is usually not as extreme as in the case of ${}^{60}$Fe (cf. Fig.~\ref{fig:EH}, top right), it remains an issue of similar magnitude as the accuracy of the energy determination and has corresponding implications for the heat production.\par
The different evaluations summarized in Fig.~\ref{fig:EH} demonstrate that in general, the radioactive heat production data for use in the geosciences have been consolidated in the last one or two decades and have improved compared with older data from the 1970s and 1980s; the clear contrast between the two subsets is a clear sign that the old data should be discarded. This consolidation notwithstanding, new measurements and improvements in the evaluations of nuclear data may still result in noticeable changes, as exemplified most strikingly by the revision of the half-life of ${}^{60}$Fe a few years ago, but also by a reevaluation of the $\beta$ energy spectra of ${}^{40}$K in progress at the time of writing, which also resulted in changes of a few per cent relative to calculations based on older values for $\langle E_\beta\rangle$.\par
The extent to which these small differences matter depends on the problem at hand. On the one hand, there are large-scale problems such as the global composition models for the Earth or other planets, or considerations of the heat budget of core dynamos. At these scales, the contents in heat-producing elements are still quite uncertain. For instance, K, Th, and U concentrations in model compositions for the Bulk Silicate Earth often differ by 10--15\%, and may in some cases even reach several tens of per cent \citep[cf.][Table~9]{Jaup:etal15}. Similarly, estimates for the content of the Earth's core especially of K vary by an order of magnitude, and thermal core evolution models that include it as one of several heat sources indicate that the effect of internal radioactive heating is of some importance mostly with regard to the initial temperature at the core--mantle boundary \citep{Nimmo15b}. In the case of the short-lived, extinct nuclides ${}^{26}$Al and ${}^{60}$Fe, the largest source of uncertainty at this point is probably the determination of initial concentrations, which can only be inferred from certain isotopic excess concentrations in their stable daughters ${}^{26}$Mg and ${}^{60}$Ni observed in primordial meteoritic material \citep{Teng17,ElSt17}, at least unless major adjustments of their half-lives are still ahead of us. For these considerations, the heating rates resulting from recent evaluations can be considered stable results, and the differences between them are presently of minor or no importance, although new constraints, e.g., provided by geoneutrino detectors \citep[e.g.,][]{Sram:etal13,Leyt:etal17} may reduce the uncertainty in the future. In this context it is worth keeping in mind that the relative amounts of the different heat-producing nuclides have changed with time due to their different half-lives, and so have their contributions: The concentrations of ${}^{40}$K, ${}^{232}$Th, ${}^{235}$U, and ${}^{238}$U have been approximately 12, 1.25, 84, and 2 times higher, respectively, at the beginning of the Solar System than today, and uncertainties in their energy output that are insignificant today would have translated into larger absolute effects in the early history of a planet.\par
On the other hand, the concentrations of these nuclides are often much better known in specific regional or local settings or for specific rock types \citep[e.g.,][]{VanSchmus95,JaMa14}, and therefore it may be possible and useful to determine the heat production of such units with a precision where even variations at the per cent level become relevant. Moreover, in rock types with high concentrations of radionuclides, in particular continental crust rock, small variations in the assumed heat production again translate into relatively large differences in absolute values. Especially in the context of enriched near-surface rocks or rock units near a rock--water interface, we may also recall from Sect.~\ref{sect:theo} that the radiation from radioactive decay has an ionizing effect. For instance, radiolysis of water and the concomitant production of H$_2$ and other gases has long been known and studied because of nuclear safety implications \citep[e.g.,][]{LeCaer11} or as an energy source for microbial lifeforms in deep-earth or astrobiological environments \citep[e.g.,][]{KPedersen97,LHLin:etal05}. In this context, it is the production rate of such radiolysis products that depends on the accessible energy from radioactive decay.

\section{Conclusions}
The heat production of the six radioactive nuclides of relevance for the thermal evolution of planetary bodies has been recalculated using recent data and accounting for the details of the decay processes to greater extent than some previous studies. The new values generally agree well with other recent studies but differ by several per cent from older published values which have been in frequent use until recently. Specifically for ${}^{40}$K, even methodically thorough earlier studies arrive at values for the decay heat that are higher by 1--2\% and would translate into a slightly higher estimate of the heat production. Even more important than the improved accuracy of the heat production calculations, however, is the awareness that the total decay energy is not fully available for heat production and that the practice of equating $Q$ with $E_H$ by ignoring the loss by neutrinos, unfortunately found in various publications, leads to a substantial overestimate of heat production for all six nuclides considered here, ranging from relatively modest 4.5--8.5\% for the three heavy nuclides to about 13 and 27\% for the short-lived ${}^{60}$Fe and ${}^{26}$Al, respectively, to as much as 97\% for ${}^{40}$K.\par
A Jupyter notebook as well as a Python script derived from it are provided as Supplementary Material to this study (\url{https://github.com/trg818/radheat}) in order to allow users to carry out calculations of their own with other values.

\subsection*{Acknowledgments}
I thank Steve Dye and three anonymous referees for a variety of helpful remarks, which helped me to clarify and improve the paper. I am also very grateful to NuDat and \url{nucleide.org} maintainers Shamsuzzoha Basunia, Jun Chen, Xavier Mougeot, Balraj Singh, and Alejandro Sonzogni for their detailed and patient answers to some questions about the data provided there, in particular issues concerning the decays of ${}^{40}$K and ${}^{210}$Bi. Initial work on the heat production of ${}^{40}$K was done with support by grant Ru 1839/1-1 from the Deutsche Forschungsgemeinschaft (DFG) to the author. The data presented in this paper are included in the text or in the supporting information, or they can be reproduced with the code provided at \url{https://github.com/trg818/radheat}.

\end{document}